\documentclass[final,5p,times,twocolumn]{elsarticle}

\usepackage{amssymb}
\usepackage{amsmath}
\usepackage{amsthm}
\usepackage{algorithm}
\usepackage{algpseudocode}
\usepackage{siunitx} 
\usepackage{booktabs} 
\usepackage{makecell} 
\usepackage{wrapfig}

\journal{Mechatronics}

\begin{document}

\begin{frontmatter}



\title{Mechanism Design Optimization through CAD-Based Bayesian Optimization and Quantified Constraints}


\affiliation[1]{organization={Department of Electromechanics, CoSys-Lab, University of Antwerp},
            city={Antwerp},
            postcode={2020},
            country={Belgium}}

\affiliation[2]{organization={AnSyMo/Cosys, Flanders Make, the strategic research centre for the manufacturing industry},
            country={Belgium}}

\affiliation[3]{organization={Department of Mathematics and Computer Science, University of Antwerp},
            city={Antwerp},
            postcode={2020},
            country={Belgium}}

\affiliation[4]{organization={Department of Computing Science and Mathematics, University of Stirling},
            city={Stirling},
            postcode={FK9 4LA},
            country={Scotland}}

\cortext[cor1]{abdelmajid.benyahya@uantwerpen.be, Groenenborgerlaan 171 2020 Antwerp Belgium}

\author[1,2]{Abdelmajid Ben Yahya\corref{cor1}}
\author[1,2]{Santiago Ramos Garces}
\author[1,2]{Nick Van Oosterwyck}
\author[3,4]{Annie Cuyt}
\author[1,2]{Stijn Derammelaere}

\begin{abstract}

This study addresses the challenge of mechanism design optimization, particularly focusing on the energy efficiency and design space of reciprocating mechanisms. The research question centers on how to effectively utilize Computer-Aided Design (CAD) simulations alongside Bayesian Optimization (BO) and a constrained design space to streamline the design optimization process, overcoming the limitations of traditional kinematic and dynamic analysis methods. The objective was to investigate and develop a novel optimization framework that integrates CAD-based simulations with a BO approach. To achieve this, the study employed a methodological approach. At first, the feasibility of a chosen mechanism design is evaluated through a sequence of CAD-motion simulations to quantify the (in)feasibility of this design. When this design appears to be feasible a CAD-based design evaluation method is started, in which the objective value is extracted by a sequence of CAD-motion simulations. In this paper, we advocate the use of non-parametric Gaussian processes to build a surrogate model of the objective function and the feasible design space constrained by static and dynamic constraints. The main research results demonstrated that the proposed CAD-based Bayesian Optimization framework could effectively identify optimal design parameters that minimize the root mean square (RMS) torque while adhering to specified static and dynamic constraints. This optimization approach significantly reduces the complexity associated with analytic methods, making it scalable to more complex mechanisms and implementable by machine builders. In conclusion, the study successfully developed a novel optimization framework that leverages CAD-based simulations and Bayesian Optimization to streamline the design process of mechanisms. The results of an emergency ventilator case study with three design parameters show a reduction of the RMS torque with 71\% after 255 CAD-based design evaluations. Moreover, the results illustrate the effectiveness of incorporating constraints into the design optimization process and the potential of this approach for achieving global optimal design in a computationally efficient manner.

\end{abstract}


\end{frontmatter}


\section{Introduction}

\label{sec:Introduction}

\begin{figure}[ht]
    \centering
    \includegraphics[width=0.99 \columnwidth]{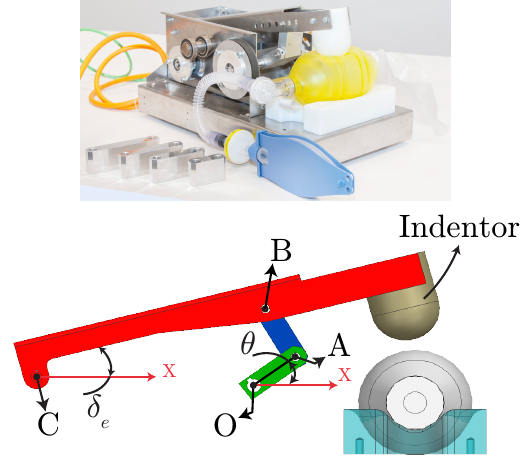}
    \caption{The use case within this research is an emergency ventilator developed by Gear Up Medical VZW \cite{Herregodts2019}.}
    \label{fig: case}
\end{figure}

Mechatronic systems account for about 70 \% of industrial energy consumption, contributing to 40-45 \% of global energy usage \cite{Waide2011}. This underscores the need for energy-saving strategies in industrial machinery, particularly through reducing losses in electric motors, which, as \cite{Saidur2010} notes, are predominantly stator losses (55-60 \%). This paper presents an optimization method to decrease energy consumption in industrial mechanisms by optimizing link lengths $|OA|$, $|AB|$, and $|BC|$, as illustrated in Figure \ref{fig: case}.


The paper demonstrates the proposed method's real-world applicability and potential through a use case, emphasizing that only the CAD model of the mechanism is needed. The use case involves an emergency ventilator, shown in Figure \ref{fig: case}, developed by a non-profit organization \cite{Herregodts2019} during the initial wave of the COVID-19 pandemic. The mechanism depicted in Figure \ref{fig: case} operates by pressing an indentor into a bag, facilitating airflow towards the patient. Figure \ref{fig: case} showcases the CAD model of the emergency ventilator, highlighting how the red beam, attached to the indentor (the end-effector), moves by rotating the input link OA around point O over an angle $\theta(t)$, which is driven by an electric motor. This ventilator was specifically designed for low- and middle-income countries, where consistent access to electricity is not guaranteed. The ventilator's design, particularly the link lengths $|OA|$, $|AB|$, and $|BC|$, is optimized for minimal energy consumption to facilitate the use of batteries.

\begin{figure*}[ht]
    \centering
    \includegraphics[width=1.8\columnwidth]{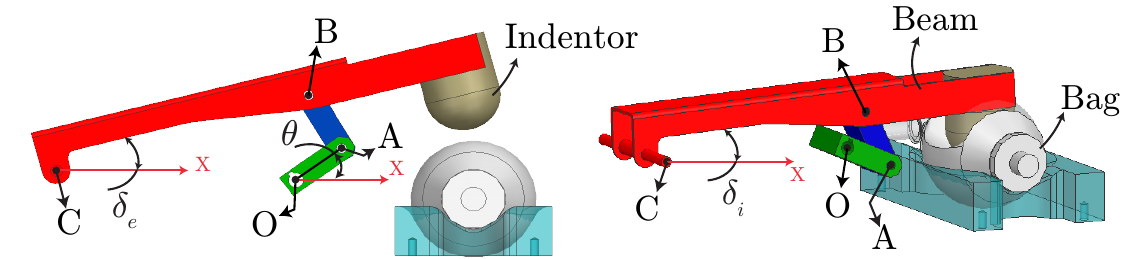}
    \caption{The mechanism shown in its ending $\delta_e$ (left) and starting position $\delta_i$ (right) of its point-to-point movement.}
    \label{fig: movement}
\end{figure*}

\begin{figure}[ht]
    \centering
    \includegraphics[width=0.99 \columnwidth]{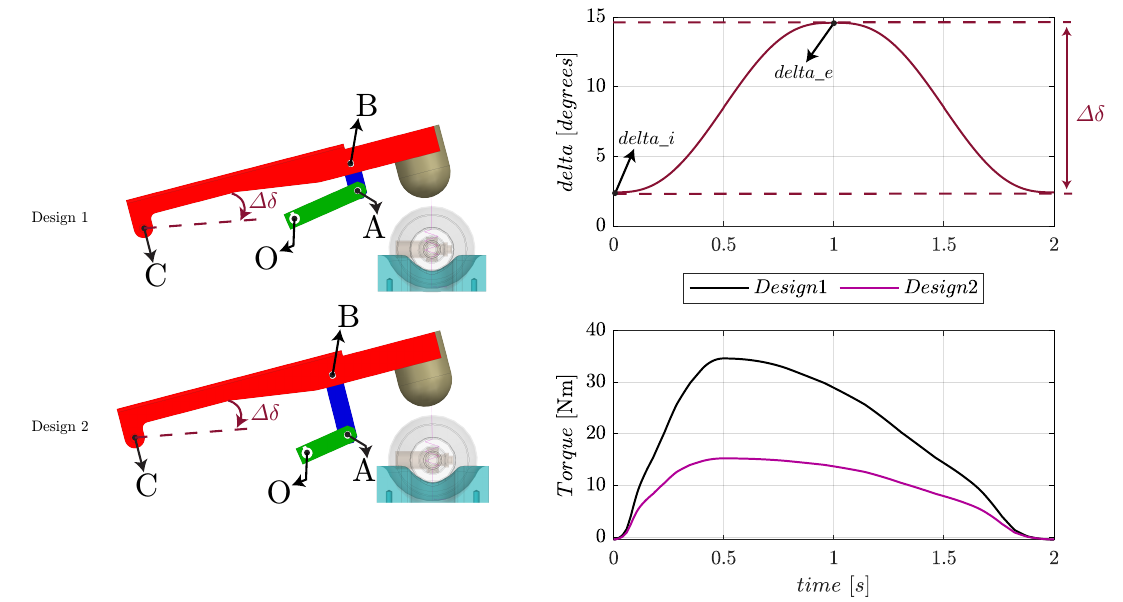}
    \caption{The emergency ventilator performs the same task, being the end-effector movement $\delta(t)$ (right), for different combinations of design parameters (DPs) $|OA|$, $|AB|$, and $|BC|$ (left). Resulting in a difference in the motor torque profile $T_m(t)$ (right).}
    \label{fig: task different lengths}
\end{figure}

The ventilator's primary function, compressing a bag via motor-driven movement $\theta(t)$, is detailed in Figure \ref{fig: movement}. The angle of $\delta_i$ and $\delta_e$ determine the mechanism's motion requirement, which represents the starting and ending positions of the indentor. Altering the link lengths $|OA|$, $|AB|$, and $|BC|$, while adhering to the motion requirements, impacts the motor's torque profiles $T_m(t)$, as shown in Figure \ref{fig: task different lengths}. These link lengths $|OA|$, $|AB|$, and $|BC|$ are thus key design parameters (DPs) in the optimization process. Prior research \cite{BenYahya2023a} has shown that such design optimization significantly reduces electric motor energy consumption by minimizing the Root Mean Square (RMS) motor torque $T_{RMS}$, which in turn reduces stator losses in the motor \cite{VanOosterwyck2022}. The significance of machine components' geometry in energy efficiency has gained increasing recognition in recent studies \cite{Sheppard2019, Carabin2017, ELKribi2013}.

This paper focuses on dimensional synthesis, specifically targeting the optimization of linkage dimensions ($|OA|$, $|AB|$, and $|BC|$) for predefined end-effector movements ($\delta_i$ to $\delta_e$), as discussed in \cite{Lee2018}. Dimensional synthesis, especially for determining precise end-effector movements in mechanisms like slider-crank and four-bar mechanisms, is extensively studied \cite{Hernandez2021a, Cabrera2002, Gogate2012, Hernandez2021}. Traditional methods involve analytic kinematic derivations, with studies like \cite{ELKribi2013, Affi2007} deriving dynamics for simple mechanisms and seeking optimal dimensions to minimize torque fluctuations. Methods for deriving the dynamics of mechanisms, such as Lagrangian dynamics and the principle of virtual work, are utilized for sensitivity analysis of dynamical systems, as discussed in \cite{Rayner2008, Dopico2014, Dopico2022}. Furthermore, principles including virtual work, vector mechanics, and Lami’s theorem \cite{Gawande2019}, along with Hamilton's principle and Lagrange multipliers \cite{Huang2009, Hsu2014, Hsu2016}, are employed in the dynamics derivation of toggle clamping mechanisms. \cite{Huang2011} demonstrates the use of analytic dynamics derivation in conjunction with genetic algorithms to optimize toggle clamping mechanism designs for enhanced mold clamping forces.

However, these analytic methods are complex, time-consuming, and prone to errors \cite{Berselli2016}. Their complexity increases with the mechanism's complexity and requires detailed component information, such as the center of gravity and mass, which change with design modifications. For example, \cite{Vanbecelaere2020} demonstrates the complexity in deriving dynamics for a mono-actuated industrial planar mechanism using the method of kinetic energy. This complexity limits the practicality of design optimization for machine builders and restricts the scalability of these dynamic equations to different or more complex mechanisms.


Recent research \cite{VanOosterwyck2019, VanOosterwyck2020} has shifted towards using CAD models to ascertain the dynamics of mechanisms, rather than relying on analytical derivation of the system's dynamics. This paper introduces a novel Computer-Aided Design (CAD)-based method for optimizing the design of industrial mechanisms. Computer-Aided Design (CAD) software, a fundamental tool for mechanical engineers \cite{Schulz2017}, is leveraged for conceptualizing mechanisms, highlighting the proposed method's industrial relevance. Unlike studies focused on Finite Element Modelling (FEM) like \cite{Le2022}, which typically require a single CAD simulation per evaluation, this study emphasizes sequential CAD motion simulations.

This study adopts a CAD-based approach for simulating the dynamics of mechanisms, incorporating key elements like volume, mass, friction, damping, and joints. It focuses on motion simulations to evaluate various design parameter configurations $|OA|$, $|AB|$, and $|BC|$ (as shown in Figure \ref{fig: task different lengths}). This method bypasses the intricate kinematic and dynamic analyses that often challenge machine builders. In a prior study by the authors \cite{BenYahya2023a}, the optimization in CAD employs heuristic and gradient-based optimizers, common in state-of-the-art techniques \cite{Hernandez2021}. However, these algorithms do not guarantee finding the global optimum. Addressing a constrained-global optimization problem effectively requires defining the feasible design space. Researchers like \cite{ELKribi2013, Affi2007, Rayner2008} often do not specify the design space, potentially leading to infeasible designs or defects in mechanism synthesis, as highlighted in \cite{Hernandez2021a}. In contrast, our earlier study \cite{BenYahya2023} established the feasible design space using kinematic analysis to facilitate the search for the global optimum.

This paper proposes a CAD-based methodology to evaluate the feasibility of design parameter combinations ($|OA|$, $|AB|$, and $|BC|$), eliminating the need for any analytic analysis. The CAD-based feasibility evaluation allows for the modeling of the feasible design space without necessitating a detailed analysis of the mechanism. This method can handle increased mechanism complexity by utilizing CAD simulations and avoiding analytical analysis, enhancing its applicability.


Incorporating CAD simulation into the optimization loop, as done in this study, inevitably increases the computational time required for solving the design problem. To address this, Bayesian optimization is employed. This stochastic process involves fitting an objective function or a surrogate model to the collected data. A critical aspect of using response surfaces for global optimization, as highlighted by \cite{Jones1998}, is the balance between exploiting the area around the currently identified minimum objective value and exploring other areas where the fitting error might be higher. Bayesian optimization can efficiently navigate towards the global optimum. Moreover, Bayesian optimization is particularly valuable as it offers a more comprehensive understanding of the optimal design, factoring in potential uncertainties. This approach allows for a more informed conclusion, suggesting that the identified optimum is likely the expected global optimum.

Bayesian optimization has seen significant advancements in recent years. Traditionally, optimization problems involving computationally intensive models or lacking gradient information have often relied on heuristic optimizers like evolutionary algorithms. These algorithms primarily depend on function evaluations, as discussed in \cite{Wang2023a}. However, a limitation of these heuristic algorithms is their inability to confirm whether the identified minimum is global, and they typically require a large number of function evaluations. The review by \cite{Greenhill2020} illustrates that recent developments in Bayesian optimization have resulted in a variety of new techniques. For instance, \cite{Sheikh2022} employed Bayesian optimization to refine the shape of a turbine for enhanced power output, utilizing Computational Fluid Dynamics (CFD) software to gather data for Gaussian processes. This approach was chosen because gradient-based algorithms often become trapped in local minima, and evolutionary algorithms demand numerous computationally expensive function evaluations to approach a minimum. Bayesian optimization, with its ability to efficiently navigate the design space and provide insights into the global optimality of solutions, presents a more effective alternative for complex optimization tasks.


The methodology presented in this paper effectively harnesses the capabilities of CAD tools and Bayesian optimization to establish a more versatile and efficient framework for optimizing the design of mechanisms, particularly complex ones. This approach is especially beneficial for achieving energy-efficient designs due to several key factors:

\begin{itemize}

    \item \textbf{Avoiding Analytics:} The use of only a CAD model for mechanism design optimization greatly simplifies the process by removing the necessity to model the mechanism's kinematics or dynamics, which are typically complex and error-prone.

    \item \textbf{CAD-Based Constraint Quantification:} This paper introduces a new method for assessing the feasibility of designs through CAD simulations, improving the accuracy and reliability of the design process. The CAD simulation produces a measurable value that reflects the feasibility or infeasibility of a design. This quantification enables the use of a surrogate model to estimate the constraint design space effectively. 

    \item \textbf{Global Optimum Search with Bayesian Optimization:} The adoption of Bayesian optimization for the search of the global optimum is a strategic choice. This stochastic algorithm identifies the optimum design and provides valuable insights into the uncertainty associated with this optimum. 
   
\end{itemize}


This paper introduces a novel approach in design optimization, applying Bayesian optimization for energy-efficient mechanism design. The methodology involves three key steps: enhancing the objective function's sampling process based on \cite{BenYahya2023a} for better robustness and efficiency (section \ref{sec:simulation}); a CAD-based method to assess the feasibility of design parameters ($|OA|$, $|AB|$, $|BC|$) (section \ref{sec:constraint}); and using Bayesian optimization to determine the global optimum, shifting from traditional methods mentioned in \cite{BenYahya2023a} (section \ref{sec:optimization}). The effectiveness of these methods is demonstrated in section \ref{sec:results} through the design optimization of an emergency ventilator. The paper concludes in section \ref{sec:conclusion}, summarizing the findings.

\section{CAD-Based Design Evaluation}

\label{sec:simulation}


This section of the paper focuses on the methodology that calculates the RMS motor torque $T_{RMS}$ through a series of CAD simulations. Multiple simulations are necessary as changes in these design parameters affect the motor's start and end angles $\theta(t)$, given the pre-defined end-effector movement $\delta_i$ to $\delta_e$, as can be seen in Figure \ref{fig: task different lengths}. The motion requirement of the mechanism during operation is defined by setting the end-effector to two specific angles: $\delta_e$, where it positions to touch the bag, and $\delta_i$, corresponding to the position for maximal compression of the bag. A kinematic transformation then determines the required motion profile  $\theta(t)$. Subsequently, this profile is used in the dynamic analysis of the mechanism to retrieve the required motor torque $T_m(t)$ and corresponding $T_{RMS}$.

\begin{figure*}[ht]
    \centering
    \includegraphics[width=1.9\columnwidth]{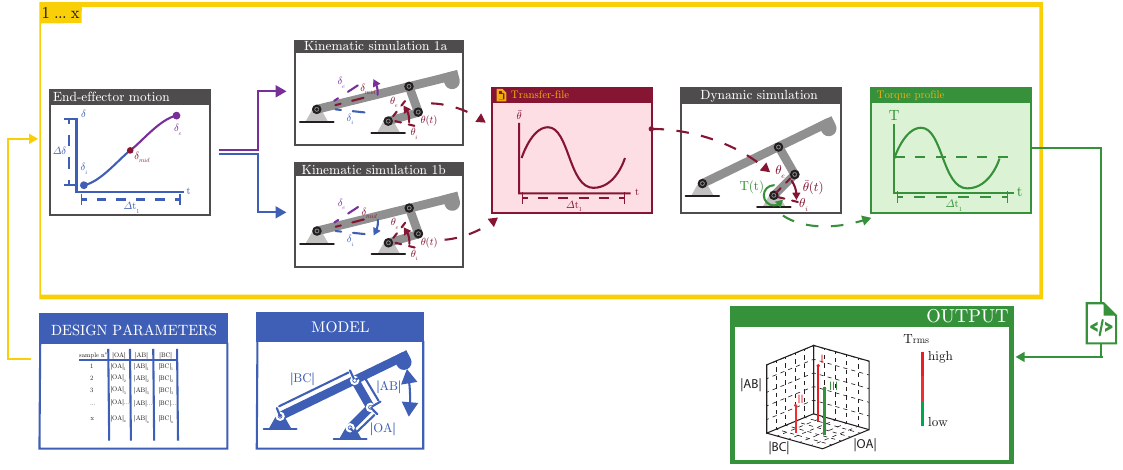}
    \caption{Three essential motion simulations and measurements are required to determine the driving torque for a specific design.}
    \label{fig: design evaluation}
\end{figure*}

The kinematic transformation mentioned is executed through kinematic simulations 1a and 1b, followed by a dynamic analysis conducted in the dynamic simulation, as depicted in Figure \ref{fig: design evaluation}. Crucially, assessing the feasibility of the design parameter combination ($|OA|$, $|AB|$, and $|BC|$) is essential before initiating these simulations. The methodology for assessing the feasibility of design parameter combinations is comprehensively described in section \ref{sec:constraint}.

The forthcoming subsections, \ref{sec:kinematic} and \ref{sec:dynamic}, are dedicated to providing a detailed description of the kinematic transformation and dynamic analysis.

\subsection{Kinematic Transformation}
\label{sec:kinematic}

\begin{figure}[ht]
    \centering
    \includegraphics[width=0.9\columnwidth]{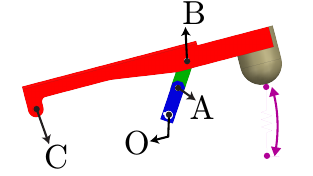}
    \caption{Choosing design parameters lengths ($|OA|$, $|AB|$, and $|BC|$) that cause bars OA and AB to align in a straight line from point B to O creates an unsolvable simulation when driving the end-effector from point C.}
    \label{fig: unsolvable motion}
\end{figure}

The initial motion simulation is designed to conduct kinematic calculations, translating the predefined end-effector movement from $\delta_i$ to $\delta_e$ into the required motor profile $\theta$. This process is divided into two simulations: "kinematic simulation 1a" and "kinematic simulation 1b," as shown in Figure \ref{fig: design evaluation}. Each simulation begins with positioning the end-effector at the midpoint of its movement ($\delta_{mid}$). From there, one simulation moves the end-effector (driven from point C) to its end position ($\delta_e$), and the other to its starting position ($\delta_i$). Splitting this kinematic transformation into two simulations tackles issues where specific designs could erroneously appear unsolvable when transitioning from $\delta_e$ to $\delta_i$. The issue is not with the design parameters ($|OA|$, $|AB|$, and $|BC|$) themselves but with the setup of the motion simulation. In designs where bars OA and AB are extended, forming a straight line between point B and the fixed point O, a singularity occurs within the motion simulation. This singularity makes it impossible to lower the end-effector by rotating the red beam at point C, as shown in Figure \ref{fig: unsolvable motion}. It is important to recognize that the aforementioned issue does not occur when the specific design is actuated from the motor, as in real-life operation. Consequently, such a design cannot be considered infeasible based on this aspect. Therefore, starting the movement from $\delta_{mid}$ ensures that this problem does not affect feasible designs.

This approach is novel compared to the previous study by the authors \cite{BenYahya2023a}, where the kinematic transformation was conducted within a single simulation. In that single simulation, moving the end-effector from $\delta_e$ to $\delta_i$ could erroneously be perceived as unsolvable in scenarios like those depicted in Figure \ref{fig: unsolvable motion}. 

\subsection{Dynamic Analysis}
\label{sec:dynamic}

As depicted in Figure \ref{fig: design evaluation}, the final simulation "Dynamic simulation" is key to determining the necessary driving motor torque $T_m(t)$ for a given set of design parameters $|OA|$, $|AB|$, and $|BC|$. 

The CAD software calculates the inertial properties of the mechanism components using their material properties. The motion simulation then formulates the equations of motion based on these inertial properties and kinematic relations, which include actual positions and their derivatives, speed, and acceleration of the individual mechanism components. Providing the motor profile $\theta(t)$ to the simulation would necessitate differentiating the measured signal to acquire both speed $\dot\theta(t)$ and acceleration $\ddot\theta(t)$. However, differentiating a signal measured within a simulation tends to amplify the inherent noise present in the numerical results. In contrast, by inputting the acceleration profile $\ddot\theta(t)$ into the simulation, the speed $\dot\theta(t)$ and motor position profile $\theta(t)$ can be obtained through integration, a process less prone to amplifying numerical simulation noise.

Hence, the kinematic transformation extracts the motor's acceleration profile $\ddot\theta(t)$ at point O from the outcomes of the two preceding simulations: Kinematic simulation 1a and Kinematic simulation 1b. This method is a departure from the previous study \cite{BenYahya2023a}, which focused on the motor profile $\theta(t)$.

The CAD software employs a numerical solver to resolve the equations of motion and determine the necessary motor torque $T_m(t)$. This eliminates the need for manual dynamic analysis by the machine builder. The dynamic simulation enables the extraction of the required motor torque $T_m(t)$ for driving the mechanism according to the predefined end-effector movement ($\delta_i$ to $\delta_e$) in kinematic simulations 1a and 1b. This facilitates calculating the $T_{RMS}$ objective value for each design, based on its specific motor torque profile $T_m(t)$.


\section{CAD-Based Feasible Design Space Quantification}

\label{sec:constraint}


The selection of design parameter combinations ($|OA|$, $|AB|$, and $|BC|$) in the methodology is determined by the optimization algorithm. Specifically, the Bayesian optimization algorithm selects new designs based on its model of the objective function. To enhance the efficiency of the optimization algorithm, it is beneficial to define the feasible design region. 

Moreover, since the objective function model tends to be lower at the boundaries of the design space, there's a likelihood that the model may continue to decrease even outside the feasible region. Without objective values from infeasible regions, the algorithm fails to learn and may persist in evaluating designs that appear to have a low objective value according to the model, yet being in fact infeasible.

To address this, it's essential to model the constraint design space for the optimization process. A basic approach might involve checking whether a simulation solves or not, but this provides limited information about the degree of infeasibility of a point, hindering the algorithm's learning about its proximity to the border of the feasible space.

Therefore, this paper introduces a method to evaluate the feasibility of a design and quantify its infeasibility. Prioritizing industrial applicability, this method relies entirely on the CAD model, extracting infeasibility quantification through CAD motion simulations. 

This method marks a significant improvement over the authors' earlier approach in \cite{BenYahya2023a}, which used one of the simulations solely for a binary feasibility check of the design parameters, offering limited insight. The new method advances beyond this binary evaluation by quantitatively assessing a design's feasibility, thereby providing more information on a design choice.


\subsection{Static constraints}
\label{sec:static con}

\begin{figure}[ht]
    \centering
    \includegraphics[width=0.9\columnwidth]{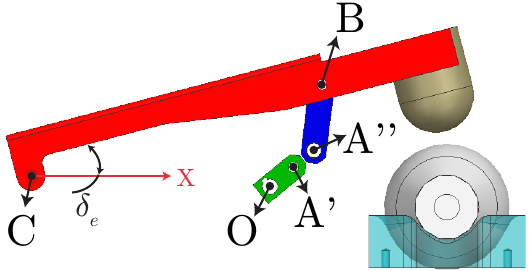}
    \caption{The combination of design parameters $|OA|$, $|AB|$, and $|BC|$ result in an infeasible design that cannot be assembled in $\delta_e$.}
    \label{fig: infeasible design}
\end{figure}

The initial step in determining feasible designs involves examining static constraints, focusing on whether a design can be assembled at the points where the end-effector is closest ($\delta_i$) and farthest ($\delta_e$) from the driver joint O. Static constraints essentially assess the assemblability of the mechanism in these critical positions \cite{BenYahya2023}. An example of a design that fails to meet these criteria is depicted in Figure \ref{fig: infeasible design}, where the chosen values for the design parameters $|OA|$, $|AB|$, and $|BC|$ result in a configuration where the link OA’ cannot connect with the link A”B.

This paper introduces a CAD-based methodology that quantifies the degree of constraint violation in an infeasible design using motion simulation.

\begin{figure}[ht]
    \centering
    \includegraphics[width=0.9\columnwidth]{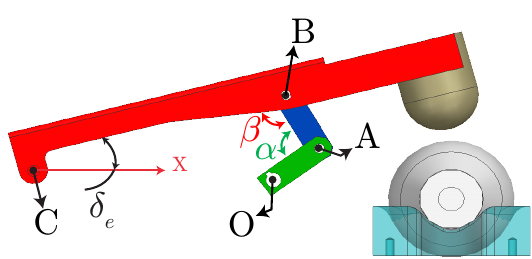}
    \caption{The relative positions between the bars are fixed through the angles $\alpha$ and $\beta$.}
    \label{fig: angles}
\end{figure}

This approach begins with a baseline design that is assemblable throughout the entire range of the end-effector movement, from $\delta_i$ to $\delta_e$. From this starting point, the relative positions of the mechanism's bars are maintained by fixing the angles between them. For instance, in Figure \ref{fig: angles}, the angles $\alpha$ and $\beta$ are constrained to remain constant at their original values, corresponding to those in the baseline design when the end-effector is at position $\delta_e$.

Once the baseline design is established with fixed angles $\alpha$ and $\beta$ between the bars, directly modifying $|OA|$, $|AB|$, and $|BC|$ is impossible due to the overconstrained nature of the CAD model—with fixed angles and all joints attached. To circumvent this, the authors suggest detaching one joint, specifically the joint at point O. This detachment allows for changes in the lengths of the bars.

\begin{figure*}[ht]
    \centering
    \includegraphics[width=1.5\columnwidth]{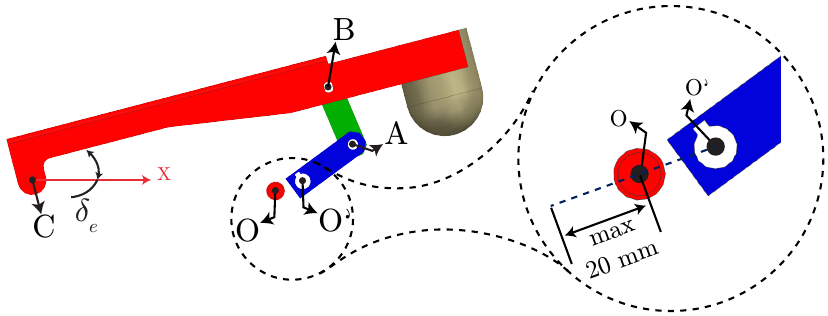}
    \caption{A combination of design parameters ($|OA|$, $|AB|$, and $|BC|$) results in a gap between points O and O' when the end-effector is in position $\delta_e$ (left). To close the gap, a motion simulation is configured to move point O' towards point O, following the straight line through these two points.}
    \label{fig: attach joint}
\end{figure*}

As depicted in Figure \ref{fig: attach joint} (left), detaching the joint at point O creates a gap between points O and O', resulting from the alteration of the design parameters $|OA|$, $|AB|$, and $|BC|$ from the baseline design. With fixed angles $\alpha$ and $\beta$, only the baseline design will align points O and O'. To assess the assemblability of a new combination of design parameters ($|OA|$, $|AB|$, and $|BC|$), a simulation is conducted to attempt closing the gap between O and O', as illustrated in Figure \ref{fig: attach joint} (right). Before starting the simulation, constraints on $\alpha$ and $\beta$ are removed, while the end-effector position is fixed at $\delta_e$. The simulation moves point O' along a straight line towards O. This simulation process is repeated with the mechanism at its starting position $\delta_i$. By conducting these simulations for both $\delta_i$ and $\delta_e$, which represent the end-effector's farthest and closest positions to the motor, the assemblability of the design parameters $|OA|$, $|AB|$, and $|BC|$ is assessed at the mechanism's extreme positions. Thus, the mechanism is considered assemblable across its entire range from $\delta_i$ to $\delta_e$ \cite{BenYahya2023}.

At this stage, the feasibility of a design can be initially assessed by checking if the motion simulation successfully aligns points O and O'. If point O' cannot align with point O the design is infeasible. However, this approach does not yet quantify the degree of infeasibility.

To address this, the authors utilize the motion simulation to track the distance between points O and O' throughout the simulation. The distance measurement's final data point, captured just before the simulation stops, provides an indication of the design's feasibility. Furthermore, the methodology extends beyond merely measuring infeasibility; it also quantifies feasibility by determining how far point O' can move past point O along the straight line connecting them. The authors have set a maximum allowable movement of point O' to be 20 mm past point O, as can be seen in Figure \ref{fig: attach joint} (right).

\begin{figure}[ht]
    \centering
    \includegraphics[width=0.9\columnwidth]{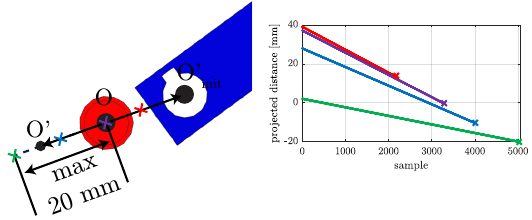}
    \caption{The final data point in the distance measurement serves as a quantifier of a design's feasibility. This value can be positive (indicated in red), signifying an infeasible design; zero (shown in purple), denoting a design that is just feasible; or negative, up to a minimum of -20 mm (represented in blue and green), indicating that the design is feasible.}
    \label{fig: constraint measurement}
\end{figure}

As depicted in Figure \ref{fig: constraint measurement}, the last data point of the distance measurement varies depending on the specific combination of design parameters ($|OA|$, $|AB|$, and $|BC|$) chosen. The relative distance of point O' to O is calculated as the projected length of the vector $\mathbf{OO'}$ onto the vector $\mathbf{OO'}_{init}$, as detailed in Equation \eqref{eq:projected length}. This approach not only identifies infeasible designs but also provides a quantitative measure of how close a design is to being feasible in both $\delta_i$ and $\delta_e$ end-effector positions.

\begin{equation}
    |OO'|\Big|_{\delta=\delta_{i},\delta_{e}} = \frac{\mathbf{OO'}_{init,\delta} \cdot (\mathbf{O} - \mathbf{O'}_{\delta})}{\sqrt{(\mathbf{O} - \mathbf{O'}_{init,\delta})_{X}^2 + (\mathbf{O} - \mathbf{O'}_{init,\delta})_{Y}^2}},
\label{eq:projected length}
\end{equation}

where, $\textbf{OO'}_{init,\delta}$ represents the vector between points O and O' at the beginning of the simulation, with the end-effector configured in either position $\delta_i$ or $\delta_e$. The vector $\textbf{O}$ indicates the fixed position of point O in the XY plane, which does not change during the simulation. $\textbf{O'}_{init,\delta}$ refers to the initial position of point O', corresponding to the end-effector in $\delta_i$ or $\delta_e$. The vector $\textbf{O'}_{\delta}$ tracks the changing position of point O' throughout the simulation, while the end-effector remains in either $\delta_i$ or $\delta_e$. The denominator in Equation \eqref{eq:projected length} is used to compute the length of $\textbf{OO'}_{init,\delta}$.

Figure \ref{fig: constraint measurement} illustrates that the feasibility quantification method produces three possible values: positive, zero, or negative. As per Equation \eqref{eq:projected length}, a positive value denotes an infeasible design, where points O and O' fail to align at the simulation's end, indicating alignment impossibility. A zero value implies that point O' aligns with point O but cannot proceed further along their connecting line, placing the design at the feasibility boundary. Conversely, a negative value indicates a feasible design where not only points O and O' can be aligned, but point O' can also move beyond this alignment, up to a maximum of 20 mm. This additional movement beyond the alignment point is incorporated to gather information about the designs on both sides of the constraint line. Consequently, this approach allows for the quantification of not only infeasible designs but also feasible designs that are near the constraint boundary, enhancing the precision of the design optimization process. 

In each design evaluation, the information obtained from the constraint quantification is utilized to construct a non-parametric surrogate model using Gaussian processes. This surrogate model serves as an approximation of the feasible design space, bounded by the static constraints, and aids in guiding the optimization process towards designs that satisfy these constraints.

\subsection{Dynamic constraints}
\label{sec:dyn con}

The static constraints outlined in chapter \ref{sec:static con} are not entirely adequate for ruling out all infeasible designs. So far, they only ensure that the mechanism can be assembled and moved from $\delta_i$ to $\delta_e$. However, it's also crucial to consider defects that may arise during the mechanism's movement. The three types of defects that can occur are order, branch, and circuit defects. The comprehensive review in \cite{Balli2002} highlights the importance of research on avoiding these defects in linkage synthesis.

As outlined in \cite{BenYahya2023}, each defect type has specific characteristics. Order defects are not relevant in this study, as only reciprocal mechanisms, which move continuously back and forth between $\delta_i$ and $\delta_e$, are considered. Branch defects occur when the transmission angle, denoted as $\beta$ in Figure \ref{fig: angles}, crosses 0 or $\pi$. This crossing results in a reversal of the motor displacement profile $\theta(t)$. Circuit defects arise when it's necessary to disassemble the linkage and reassemble it in another circuit to complete its motion. This also involves the transmission angle $\beta$ passing through 0 or $\pi$, leading to a reversal in $\theta(t)$. Such reversals are problematic in this context, as the movement is driven by a single joint at point O. A direction change in the motor displacement profile $\theta(t)$ indicates that the mechanism passes a transmission angle of 0 or $\pi$, a situation where connected links become collinear. This collinearity can cause excessively high torques, hindering movement from $\delta_i$ to $\delta_e$.

In summary, ensuring that the motor displacement profile $\theta(t)$ remains monotonic throughout the end-effector's movement from $\delta_i$ to $\delta_e$ is crucial to eliminate potential circuit and branch defects.

The evaluation of dynamic constraints relies on the availability of the motor displacement profile $\theta(t)$. This motor profile, necessary for moving the designed mechanism from $\delta_i$ to $\delta_e$, can only be obtained after completing kinematic simulations 1a and 1b. These simulations provide the required $\theta(t)$ profile. Additionally, the motor speed profile $\dot\theta(t)$, which is also essential for this analysis, can be extracted following these simulations.

As detailed in Algorithm \ref{alg: dyn con}, the motor speed profile $\dot\theta(t)$ is crucial for identifying changes in the direction of the motor displacement profile $\theta(t)$. A reversal in the direction of $\theta(t)$ is signaled by a sign change in $\dot\theta(t)$. When such a sign change is detected, the dynamic constraint is assigned a nonzero value. This value specifically represents the extent of motor displacement during the interval where the mechanism violates the constraint.

\begin{figure}[ht]
    \centering
    \includegraphics[width=0.9\columnwidth]{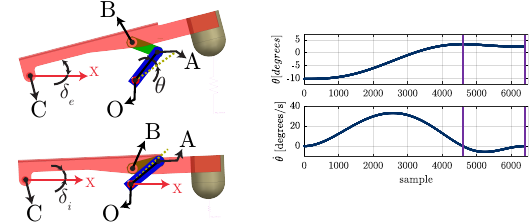}
    \caption{Design parameters ($|OA|$, $|AB|$, and $|BC|$) lead to the left design, assemblable in $\delta_e$ and $\delta_i$ but with a speed profile sign change. Dynamic constraints assess the complete movement, quantifying infeasibility by the $\theta$ range with speed sign change, shown between purple vertical lines on the right.}
    \label{fig: dyn constraint}
\end{figure}

Figure \ref{fig: dyn constraint} demonstrates an instance of dynamic constraint violation, showcasing the corresponding motor displacement ($\theta$) and speed ($\dot\theta$) profiles. In Figure \ref{fig: dyn constraint}, the occurrence of a sign change in the speed profile, detected in line 3 in Algorithm \ref{alg: dyn con}, indicates a violation of the dynamic constraint. Quantifying the dynamic constraint's violation is done by the range of motor displacement during which the speed profile $\dot\theta(t)$ deviates in sign from a predetermined reference.


\begin{algorithm}[ht]
\caption{Dynamic Constraint Calculation}
\label{alg: dyn con}
\begin{algorithmic}[1]

\If{\textbf{all}($\dot\theta \geq 0$) \textbf{or} \textbf{all}($\dot\theta \leq 0$)}
    \State $dynamic\_constraint \gets 0$

\ElsIf{\textbf{all}($\dot\theta_{bot} \geq 0$) \textbf{or} \textbf{all}($\dot\theta_{top} \leq 0$)}
    \State $dynamic\_constraint \gets \textbf{range}(\theta(\dot\theta < 0))$
\Else
    \State $dynamic\_constraint \gets \textbf{range}(\theta(\dot\theta > 0))$
\EndIf
\end{algorithmic}
\end{algorithm}

For each design evaluation, a specific value is assigned to the dynamic constraint, corresponding to the design being assessed. This value is then utilized to develop a non-parametric surrogate model using Gaussian processes. This surrogate model of the dynamic constraint provides an approximation of the design space that is limited by the dynamic constraint. 

\begin{figure}[ht!]
    \centering
    \includegraphics[width=0.9\columnwidth]{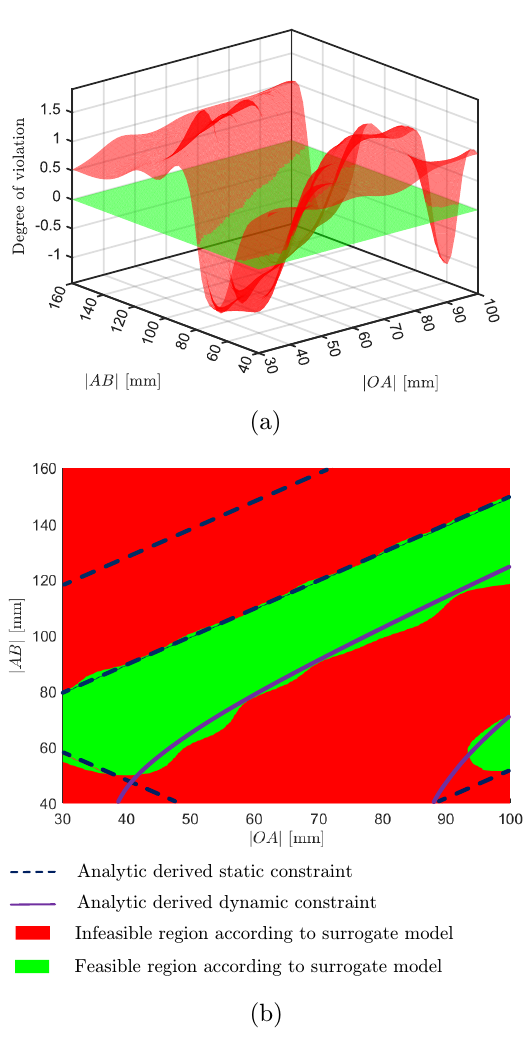}
    \caption{The surrogate model in (a) delineates the feasible design space for two parameters, $|OA|$ and $|AB|$, with a red surface indicating feasible areas (below the green plane) and probable infeasible zones (above the green plane). In (b), the model incorporates overlays of analytically derived static (dashed blue lines) and dynamic (purple lines) constraints, identifying the same design space's feasible regions in green.}
    \label{fig: constraint model}
\end{figure}

Figure \ref{fig: constraint model} (a) illustrates the surrogate model that approximates the feasible design space based on two design parameters ($|OA|$ and $|AB|$) for demonstration purposes. The red surface indicates a design's degree of violation by a value between -1 and 1, mapping out the feasible design space. Designs with violation values exceeding 0 (above the green plane) are likely infeasible, whereas those less than 0 (below the green plane) are likely feasible. In Figure \ref{fig: constraint model} (b), the surrogate constraint model is overlaid with the design space's analytically derived static and dynamic constraints, as detailed in \cite{BenYahya2023}. This combination, shown in a 2D overlay, confirms that regions enclosed by the analytically derived static (dashed blue lines) and dynamic (purple lines) constraints match the feasible (green) areas identified by the surrogate model. The comparison reveals that the surrogate model closely approximates the analytic constraints within the design space, demonstrating its effectiveness.

The model of the design space as shown in Figure \ref{fig: constraint model} (a), constrained by both dynamic and static constraints, aids the optimization process. The approach guides the selection of new combinations of design parameters ($|OA|$, $|AB|$, and $|BC|$) towards a better resulting objective value and a design with a high probability of feasibility for evaluation.

\section{Optimization Approach}

\label{sec:optimization}

In this work, the design optimization problem, as stated in Equation \eqref{eq:objective}, is to find the optimal design (being lengths $|OA|$, $|AB|$, and $|BC|$) leading to a minimal $T_{RMS}$ for this mechanism. 

\begin{equation}
\begin{array}{ll}
\min : & T_{RMS}(\textbf{x}) \\
\text { subject to: } & \text{Static constraint(\textbf{x}) $\leq$ 0} \\
& \text{Dynamic constraint(\textbf{x}) $=$ 0} \\
& x_i \in\left[x_{i \min }, x_{i \max }\right] ; \quad x_i \in \textbf{x}
\end{array}
\label{eq:objective}
\end{equation}
where
$$
\begin{array}{ll}
     &T_{RMS} \ \text{is the objective function,}  \\
     &\text{The Static and Dynamic constraint evaluates the feasibility} \\
     &\text{of a certain design,} \\
     &\textbf{x} \ \text{is a vector, which contains the independent} \\
     &\text{design parameters} \ \vert OA \vert, \vert AB \vert \ \text{and} \ \vert BC \vert, \\
     &x_{i \min } \ \text{and} \ x_{i \max } \text{define the limits of each design variable} \ x_i.
\end{array}
$$

In a previous study \cite{BenYahya2023a}, the authors addressed the optimization problem using two prevalent algorithms in design optimization \cite{Hernandez2021a}, Sequential Quadratic Programming (SQP) and Genetic Algorithm (GA). The findings from \cite{BenYahya2023a} reveal that while the GA minimizes the likelihood of getting stuck in a local minimum, it does not guarantee to find the global optimum \cite{Bodenhofer1999a}. Moreover, GA is computationally expensive due to the high number of design evaluations required to identify an optimal solution. Conversely, the outcome of the SQP algorithm is heavily dependent on the chosen initial point. Starting from an initial design, SQP progresses along the steepest negative gradient towards a minimum, leading to quicker convergence. Thus, this method significantly increases the risk of getting stuck in a local optimum.

\begin{figure}[ht]
    \centering
    \includegraphics[width=0.9\columnwidth]{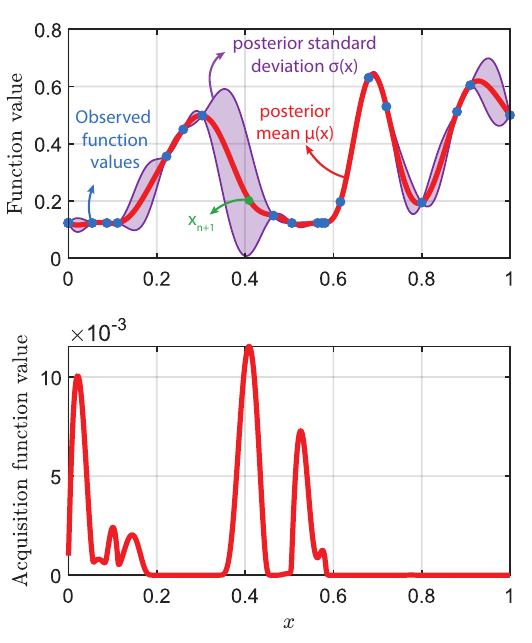}
    \caption{A trained Gaussian Process by observed function value points (blue), resulting in a function model with a posterior mean ($\mu$) depicted in red and a posterior standard deviation ($\sigma$) shown in purple. Below the surrogate model, the acquisition function utilized by BO is constructed, highlighting the new expected and feasible optimum $\textbf{x}_{n+1}$, identified as the x-value with the highest acquisition function value.}
    \label{fig: GP_graph}
\end{figure}

To tackle the challenge of uncertainty in reaching the global optimum and to decrease the extensive number of design evaluations, the authors adopt Bayesian Optimization (BO) as an approach. Bayesian optimization employs a stochastic surrogate model to approximate an expensive objective function and its constraints based on a limited set of observed function values. This model, trained using the observed function value points, results in a surrogate model with a posterior mean ($\mu$) and a posterior standard deviation ($\sigma$) representing uncertainty, as depicted in 1D for illustration at the top of Figure \ref{fig: GP_graph}.

These surrogate models are constructed using Gaussian Processes (GP), which extend Gaussian distributions to function spaces \cite{Rasmussen2006}. A GP is a distribution over functions completely defined by its mean function $\mu$ and covariance function (or kernel) $k$, as stated in Equation \eqref{eq: GP}. 

\begin{equation}
     f(\textbf{x}) \sim \mathcal{N}(\mu,\,k(\textbf{x,x'}))
     \label{eq: GP}
\end{equation}

Here, $\mathcal{N}$ denotes the normal distribution, and $k(\textbf{x,x'})$ is calculated using the squared-exponential covariance function, as indicated in Equation \eqref{eq: COV}:

\begin{equation}
     k(\textbf{x,x'}) = \sigma^{2}\exp(-\frac{(\textbf{x}-\textbf{x}')^{2}}{2\textbf{l}^{2}}).
\label{eq: COV}
\end{equation}

The covariance function provides the correlation between points \textbf{x} and \textbf{x'} in the design space, parameterized by the amplitude parameter $\sigma$ and the length scale $\textbf{l}$, known as hyperparameters. Note that the length scale $\textbf{l}$ is a diagonal matrix with a dimension corresponding to the number of design parameters. Given $n$ observations of the objective function $T_{RMS}(\textbf{x})$ at points $\textbf{x}_i$, the complete covariance/kernel matrix is computed as in Equation \eqref{eq: Matrix}.

\begin{equation}
    \mathbf{K}=\left[\begin{array}{ccc}
    k\left(\mathbf{x}_1, \mathbf{x}_1\right) & \ldots & k\left(\mathbf{x}_1, \mathbf{x}_n\right) \\
    \vdots & \ddots & \vdots \\
    k\left(\mathbf{x}_n, \mathbf{x}_1\right) & \ldots & k\left(\mathbf{x}_n, \mathbf{x}_n\right)
    \end{array}\right]
    \label{eq: Matrix}
\end{equation}

As outlined by \cite{Shende2021}, Bayesian Optimization (BO) comprises two primary components. The first is the probabilistic surrogate models for the objective and constraint models, which mimic the behavior of, computationally expensive, functions as shown at the top of Figure \ref{fig: GP_graph}. The second involves using an acquisition function based on these probabilistic models, guiding the selection of the next optimal evaluation point. This acquisition function calculates a value at any unobserved point $\textbf{x}{n+1}$, based on the posterior distribution that provides a posterior mean ($\mu$) and standard deviation ($\sigma$) at $\textbf{x}{n+1}$ \cite{Rasmussen2006}, leveraging the Gaussian Process (GP) properties. In this study, we employ the Expected Improvement (EI) method, an improvement-based acquisition function introduced by \cite{Jones1998}. The EI acquisition function assesses the surrogate model's predicted mean against the present optimal minimum objective value, incorporating the standard deviation to quantify the anticipated improvement at any point within the design space. Additionally, this paper advocates to incorporate the static and dynamic constraints, discussed in Section \ref{sec:constraint}, by a constraint surrogate model to predict the probability of feasibility at any point in the design space. This probability is integrated into the expected improvement, leading to the use of the constrained expected improvement (cEI) as the acquisition function. The acquisition function, depicted at the bottom of Figure \ref{fig: GP_graph}, identifies the new expected and feasible optimum $\textbf{x}_{n+1}$ as the x-value with the highest ordinate. Each iteration assesses the computationally costly objective function at this new optimum $\textbf{x}_{n+1}$. BO employs cEI to navigate within the design space, ensuring that not only a certain area is exploited but also that exploration of the design space is performed to avoid sub-optimal results and aim for the global optimum.

\begin{figure}[ht]
    \centering
    \includegraphics[width=0.9\columnwidth]{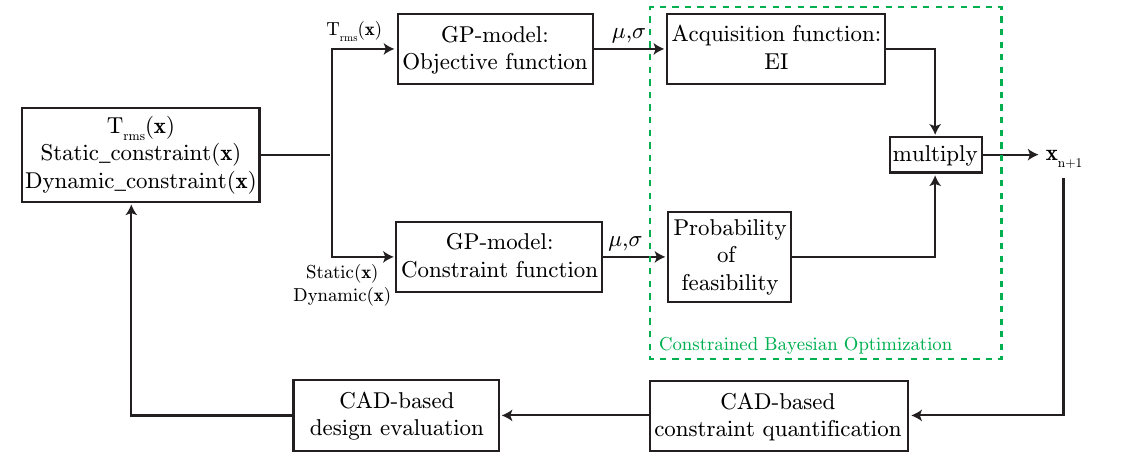}
    \caption{The entire optimization workflow, integrating Gaussian Process (GP) models for both objective and constraint functions. Utilizing these surrogate models, the workflow initiates constrained Bayesian optimization, which establishes the acquisition function and selects an unobserved point $\textbf{x}_{n+1}$ as the expected and feasible optimum. At this point in the design space, the constraint and objective values are ascertained through CAD motion simulations, the results of which are fed back into the GP model to further refine the surrogate model.}
    \label{fig: BO_workflow}
\end{figure}

Figure \ref{fig: BO_workflow} presents the complete optimization process. Initially, a point of interest is selected in the design space, for which the objective $T_{RMS}(\textbf{x})$, Static constraint(\textbf{x}), and Dynamic constraint(\textbf{x}) value are determined. The constraint values are established using the CAD-based constraint quantification method detailed in Section \ref{sec:constraint}, while the objective value is derived from the CAD-based design evaluation approach described in Section \ref{sec:simulation}. These values for the objective and constraint functions serve as training data for the Gaussian Process models of the objective and constraint functions. The constrained bayesian optimization process employs the posterior distribution to obtain a posterior mean ($\mu$) as the predicted objective and constrained value and a posterior standard deviation ($\sigma$), from these surrogate models. These posterior parameters are used to formulate the acquisition function as in \cite{Gelbart2014}, which then guides the selection of an unobserved point as the expected and feasible optimum. This BO approach provides insights into the uncertainty associated with the obtained optimum. This aspect of uncertainty is pivotal, as it allows for the establishment of a threshold. Users can set this threshold to determine that any further improvements in the optimum will not exceed the predefined threshold, thereby suggesting that the attained optimum is, within a certain level of uncertainty and across all evaluated points, likely the global optimum.

\section{Results \& Discussion}

\label{sec:results}

In this section, the authors discuss the effectiveness of Bayesian Optimization (BO) in the design optimization of an emergency ventilator. The results of the BO method, labeled as BO$_{Con1}$ and BO$_{Con2}$, are presented in Table \ref{tab:designoptimization}. This is compared with the outcome from the Sequential Quadratic Programming (SQP) optimization, which incorporates design feasibility as a nonlinear inequality constraint and is listed in the same table as SQP$_{Con1}$ and SQP$_{Con2}$. The study also contrasts these findings with previous research outcomes. Specifically, in \cite{BenYahya2023a}, both SQP and Genetic Algorithm (GA) were utilized without the integration of CAD-based constraint quantification or insights into the feasible design space, relying solely on simulations for a binary feasibility assessment of design parameters. These earlier results are presented in Table \ref{tab:designoptimization} under the labels SQP and GA.

In contrast, \cite{BenYahya2023} adopted an analytical method to define the feasible design space, facilitating the use of the sparse interpolation method. This approach involved modeling the objective function with the fewest possible samples and then determining the optimal objective value within the convex hull of the feasible design space using a brute-force method. The result is displayed in Table \ref{tab:designoptimization} under the label SI. The limitation of this SI method is that it guarantees finding the global optimum only within the convex hull that spans the larger part of the feasible design space, thereby excluding smaller regions that might contain the global optimum.

\begin{table}[ht]
\centering
\caption{Saving potential achieved by design optimization.}
\label{tab:designoptimization}
\setlength\extrarowheight{5pt}
\resizebox{0.48\textwidth}{!}{%
\begin{tabular}{
    l
    S[table-format=2.2]
    S[table-format=3.2]
    S[table-format=3.2]
    S[table-format=2.2]
    S[table-format=2.2]
    S[table-format=2.1]
    S[table-format=2.1]
    S[table-format=3.0]
}
\toprule
{Design} & {\makecell{$\vert OA \vert$ \\ (\si{\milli\meter})}} & {\makecell{$\vert AB \vert$ \\ (\si{\milli\meter})}} & {\makecell{$\vert BC \vert$ \\ (\si{\milli\meter})}} & {\makecell{$T_{\text{rms}}$ \\ (\si{\newton\meter})}} & {\makecell{$T_{\text{max}}$ \\ (\si{\newton\meter})}} & {\makecell{$T_{\text{rms}}$ \\savings \\(\%)}} & {\makecell{$T_{\text{max}}$ \\savings \\(\%)}} & {\makecell{Number \\of \\ evaluations}} \\

\midrule
Original & 53    & 65    & 282   & 7.9  & 13.3 & {-}   & {-}   & {-}   \\
GA \cite{BenYahya2023a}       & 82.68 & 141.25 & 281.8 & 3.1  & 5.4  & 60  & 59  & 399   \\
SQP \cite{BenYahya2023a}      & 30    & 76.22  & 271.75 & 3.4  & 5.2  & 57  & 61  & 39    \\
SI \cite{BenYahya2023}       & 33.2    & 79.4  & 266.1 & 2.6  & 4.4   & 67  & 67   & 618    \\
SQP$_{Con1}$  & 30.02    & 79.2  & 269.39 & 2.3  & 4.1   & 71  & 69  & 64   \\
SQP$_{Con2}$  & 35.84    & 106  & 295.59 & 2.7  & 4.7  & 66  & 65  & 107   \\
BO$_{Con1}$   & 30.21    & 72.36  & 258.4 & 2.3  & 4.1  & 71  & 69  & 255   \\
BO$_{Con2}$   & 30.43  & 80.38  & 270.55 & 2.3  & 4.1   & 71  & 69  & 363   \\
\bottomrule
\end{tabular}%
}
\end{table}

The data in Table \ref{tab:designoptimization} generally indicates that incorporating constraints into the design optimization process improves the quality of the obtained optimum. The SQP algorithm, enhanced by the inclusion of the constraint quantification method and a well-defined finite difference step size, achieved an optimum of 2.3 Nm after 64 iterations (SQP$_{Con1}$).This point of optimum is defined by the criterion that the objective value does not decrease by more than 0.001 Nm over 3 consecutive objective value reductions. Furthermore, initiating the SQP algorithm with constraints from a different initial point ($|OA|$=95, $|AB|$=60, $|BC|$=265) led to an optimum of 2.7 Nm after 107 iterations, as indicated by SQP$_{Con2}$, adhering to the same minimal objective value decrease criterion and identical algorithmic settings. This variation in results based on the starting point reinforces the understanding that outcomes of gradient-based optimization are highly sensitive to initial conditions.

Regarding Bayesian Optimization, both results showed an objective value of 2.3 Nm, equating to 71\% savings in the objective $T_{RMS}$. However, the iterations varied: BO$_{Con1}$ reached the optimum in 255 iterations, while BO$_{Con2}$, starting from the same initial conditions as SQP$_{Con2}$, required 363 iterations. The number of iterations is determined based on the stopping criterion, as set for the SQP algorithm, that the objective value cannot improve more than 0.001 Nm over 3 consecutive iterations. Moreover, as Gaussian processes provide insights into the uncertainty associated with the obtained optimum, one can calculate this uncertainty based on the attained objective function surrogate model at iteration 255 and 363 while using a 95\% standard deviation. Given both Gaussian processes models we've developed from different starting points with respectively 255 and 363 iterations, we consider all points in our design space, marked by a granularity of 1 mm, for which the constraint model is 95\% sure that it is feasible. For each of these points, the obtained objective function model holds 95\% confidence that the function value cannot be decreased. Based on this model's insights, we're led to believe that the optimum we've identified stands as the global optimum.


\section{Conclusion}

\label{sec:conclusion}

This paper presents a comprehensive study on the design optimization of an emergency ventilator, emphasizing energy efficiency through the use of Computer-Aided Design (CAD) and Bayesian Optimization (BO). The ventilator, initially developed for low- and middle-income countries during the COVID-19 pandemic, required optimization to minimize electric energy consumption. The study optimizes geometric design parameters ($|OA|$, $|AB|$, and $|BC|$) towards a minimal Root Mean Square (RMS) motor torque, directly linked to energy consumption.

The authors propose a novel approach that utilizes CAD motion simulations to simplify the optimization process, avoiding the complexities of traditional kinematic and dynamic analyses, which are generally less general and more cumbersome. The method described in this paper evaluates different design parameter combinations ($|OA|$, $|AB|$, and $|BC|$) and quantifies the feasibility, facilitating the optimization process. By employing CAD-based feasibility evaluation, the method facilitates delineating the feasible design space without needing the detailed mechanism analysis typically found in the literature, which is cumbersome and demands user expertise. This optimization approach can handle increased mechanism complexity by utilizing CAD simulations and avoiding analytical analysis, enhancing its applicability.

Bayesian Optimization is introduced to overcome the impracticality of brute-force approaches in seeking global optimum designs. BO employs probabilistic surrogate models based on Gaussian Processes, efficiently navigating the design space and balancing exploration and exploitation. This approach also provides insights into the uncertainty of the optimum, enhancing confidence in the attained optimum being the global optimum.

The study contrasts the Bayesian Optimization (BO) method, achieving a root mean square torque ($T_{RMS}$) of 2.3 Nm in 255 or 363 iterations, against the traditional Sequential Quadratic Programming (SQP) method under constraints, which attains the same objective value of 2.3 Nm in 64 iterations or 2.7 Nm in 107 iterations, varying by the algorithm's starting position. Additionally, it presents comparison results with a binary feasibility check of the design parameters, where SQP reaches an objective value of 3.4 Nm in 39 iterations, and the Genetic Algorithm (GA) achieves 3.1 Nm after 399 iterations. These outcomes underscore the initial condition sensitivity of gradient-based optimizations and the GA's computational intensity. Besides that, the findings suggest that integrating constraints overall enhances the optimization's effectiveness. The study concludes that the BO method, with its ability to predict the objective value at new points and assess the feasibility, effectively reduces the evaluation of infeasible designs, leading to a more efficient optimization process and potentially achieving the global optimum.

\section*{CRediT authorship contribution statement}
\textbf{Abdelmajid Ben Yahya:} Conceptualization, methodology, software, formal analysis, investigation, data curation, writing—original draft, visualization. \textbf{Santiago Ramos:} Conceptualization, methodology, writing—review and editing. \textbf{Nick Van Oosterwyck:} Conceptualization, methodology, writing—review and editing. \textbf{Annie Cuyt:} Conceptualization, writing—review and editing, supervision. \textbf{Stijn Derammelaere:} Conceptualization, methodology, writing—review and editing, supervision.

\section*{Declaration of competing interest}
The authors declare that they have no known competing financial interests or personal relationships that could have appeared to influence the work reported in this paper.

\section*{Acknowledgment}
This research received no external funding.

\section*{Data availability}
The data presented in this study are available on request from the corresponding author.

\section*{Declaration of Generative AI and AI-assisted technologies in the writing process}
During the preparation of this work the authors used ChatGPT-4 in order to suggest structure for individual sentences and paragraphs. After using this tool/service, the authors reviewed and edited the content thoroughly and take full responsibility for the content of the publication.

\appendix

 \bibliographystyle{elsarticle-num} 
 \bibliography{References}






\section*{Short Biography of Authors}

\begin{wrapfigure}{l}{25mm} 
    \includegraphics[width=1in,height=1.25in,clip,keepaspectratio]{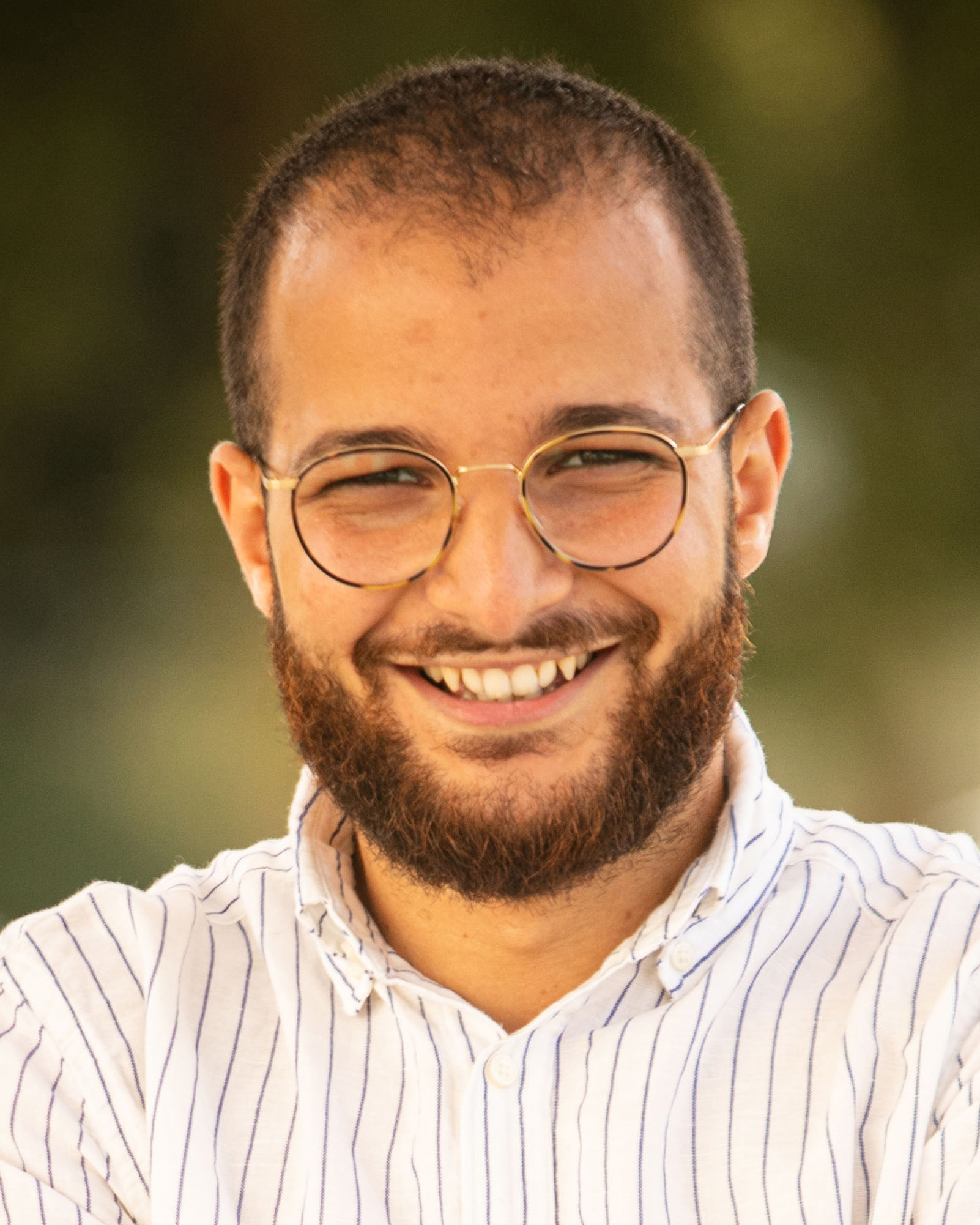}
\end{wrapfigure}\par
  \textbf{Abdelmajid Ben yahya}, born in 1996 in Antwerp, Belgium, is an engineer and researcher. In 2019, he received his Master of Science degree in Electromechanical Engineering Technology from the University of Antwerp, where he graduated summa cum laude. Following his graduation, he was employed as a Project Engineer at the University of Antwerp, working on technology transfer to small and medium-sized enterprises in Flanders. Currently, Ben Yahya is a PhD student at the University of Antwerp, starting from 2021 in the Cosys-Lab research group, researching the optimization of design for reciprocating mechatronic systems.\par
\vspace{1mm}
\begin{wrapfigure}{l}{25mm} 
    \includegraphics[width=1in,height=1.25in,clip,keepaspectratio]{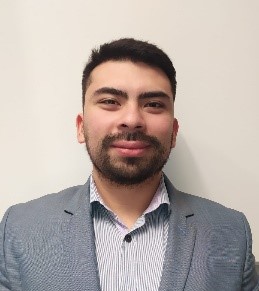}
\end{wrapfigure}\par
  \textbf{Santiago Ramos}was born in 1992 in Popayan, Colombia. He earned a Joint Master's in Mechatronic Engineering (EU4M) in 2020 from the University of Oviedo. After graduation, he became a researcher in motion control at the University of Antwerp. Since 2022, Santiago has been a PhD researcher at the University of Antwerp in the Cosys-Lab research group, collaborating with the Belgian Nuclear Research Center (SCK-CEN). His research focuses on optimizing and controlling the Isotope Separator OnLine technique (ISOL).\par
\vspace{1mm}
\begin{wrapfigure}{l}{25mm} 
    \includegraphics[width=1in,height=1.25in,clip,keepaspectratio]{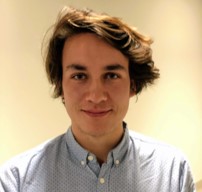}
\end{wrapfigure}\par
  \textbf{Nick Van Oosterwyck} was born in 1996 in Antwerp, Belgium. He received his MSc degree (magna cum laude) in Electromechanical Engineering Technology from the University of Antwerp in 2018 after which he started as a Ph.D. researcher at the Cosys-Lab research group. In 2019, he received additional funding from the FWO for his PhD project on energy-optimal motion profile optimization, which he successfully defended in 2023. Since then, he has been working as a teaching assistant within the Department of Electromechanics at the University of Antwerp where he is responsible for various seminars and lab sessions  focused on Motion Control and Electrical Engineering, as well as integrating the research findings into the educational curriculum. His research interests include CAD motion simulations, motion profile optimization and optimization algorithms in general.\par
  
\newpage

\begin{wrapfigure}{l}{25mm} 
    \includegraphics[width=1in,height=1.25in,clip,keepaspectratio]{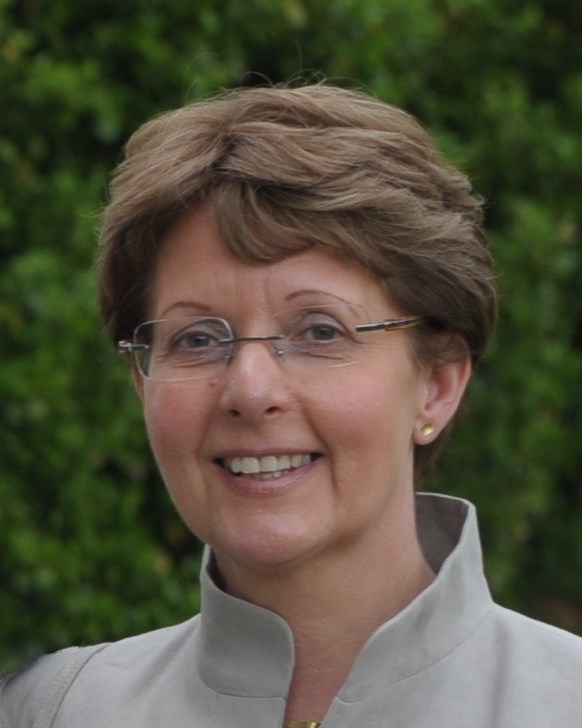}
\end{wrapfigure}\par
  \textbf{Annie Cuyt} is an emerita full professor at the University of Antwerp's Faculty of Science. She received her Doctor Scientiae degree in 1982 summa cum laude and with the felicitations of the jury. She was a research fellow with the Alexander von Humboldt Foundation and honored with a Masuda Research Grant. Cuyt is a lifetime member of the Royal Flemish Academy of Belgium for Sciences and Arts and author of over 220 peer-reviewed publications, books, and organizer of international events. Her research focus is in numerical approximation theory and its applications in scientific computing, computer science, engineering, and bio-informatics, specifically rational approximation and its relation to sparse interpolation and exponential analysis. Cuyt has served on several national and international science foundation boards and prestigious international award juries. In 2005, she helped establish a HPC center for Flanders which has grown into a successful ongoing project.\par
\vspace{1mm}
\begin{wrapfigure}{l}{25mm} 
    \includegraphics[width=1in,height=1.25in,clip,keepaspectratio]{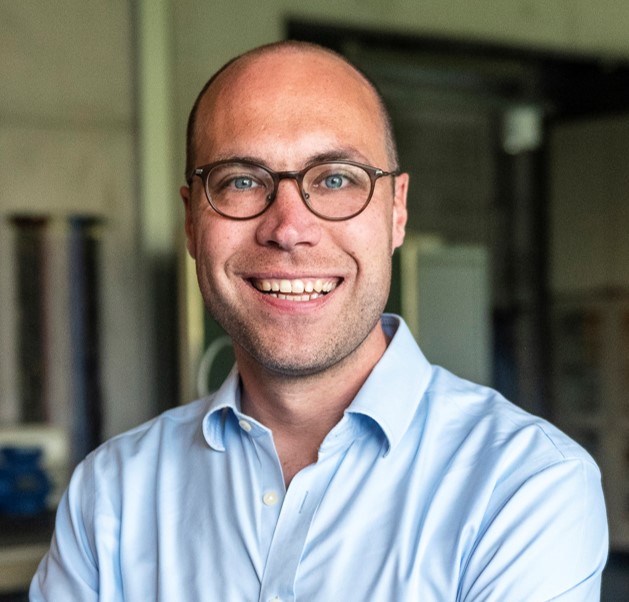}
\end{wrapfigure}\par
  \textbf{Stijn Derammelaere}, born in Kortrijk, Belgium, in 1984, earned his Master's in automation in 2006 from the Technical University College of West-Flanders, Belgium. He completed his PhD in 2013 at Ghent University, Belgium, focusing on control engineering and mechatronic systems co-design. Since 2017, he continues this research at the University of Antwerp, now serving as an associate professor in mechatronics. His expertise includes co-design, motion control, and optimization of mechatronic systems. He is the associate editor of Discover Mechanical Engineering Springer journal and teaches control engineering and motion optimization at the faculty of applied engineering.\par

\end{document}